\documentclass[prd,twocolumn,amsmath,amsfonts,amssymb,showpacs]{revtex4} 
\usepackage{graphicx}
\usepackage{mathptmx}      
\usepackage{latexsym}
\def\beq{\begin{equation}}
\def\eeq{\end{equation}}
\def\bea{\begin{eqnarray}}
\def\eea{\end{eqnarray}}

\begin{document}

\title{Corrected entropy of the rotating black hole solution
of the new massive gravity using the tunneling method and Cardy formula }

\author{Behrouz Mirza}
\email{b.mirza@cc.iut.ac.ir}
\affiliation{Department of Physics, Isfahan University of Technology, Isfahan 84156-83111, Iran}

\author{Zeinab Sherkatghanad}
\email{z.sherkat@ph.iut.ac.ir}
\affiliation{Department of Physics, Isfahan University of Technology, Isfahan 84156-83111, Iran}



\begin{abstract}
We study the AdS rotating black hole solution for the
Bergshoeff-Hohm-Townsend (BHT) massive gravity in three
dimensions. The field equations of the asymptotically  AdS black
hole of the static metric can be expressed as the first law of
thermodynamics, i.e. $dE=TdS-PdV$. The corrected Hawking-like
temperature and entropy of asymptotically AdS rotating black hole
are calculated using the Cardy formula and the tunneling method. Comparison of these methods will help
identify the unknown leading correction parameter
$\beta_1$ in the tunneling method.

\end{abstract}
\maketitle

\section{Introduction}
As shown by Hawking, black holes are not really black but emit all kinds of particles with a  perfect black body spectrum
\cite{1,2,3,4}. The Hawking black hole temperature is proportional
to its surface gravity and its entropy is proportional to its
horizon area. This is known as the celebrated Bekenstein-Hawking
area law $S_{BH}=\frac{A}{4}$. There are  several different
methods for calculating  the correction to the semiclassical
Bekenstein-Hawking entropy. These are based on statistical
mechanical arguments, field theory methods, quantum geometry,
Cardy formula, tunneling method, etc. (The corresponding
literature is rather extensive; for a partial selection, see Refs.
\cite{5,6,7,8,LOV,BTZ}). Among these, the tunneling method is
based on two variants called null geodesic method
\cite{tunneling1} and Hamiltonian-Jacobi method \cite{tunneling2}.

Bergshoeff, Hohm and Townsend (BHT)  recently advanced a theory
of massive gravity in three dimensions with remarkable properties
\cite{TOW}. The BHT theory appears to be unitary  and
renormalizable  and several exact solutions have been found \cite
{ {UN}, {RE}}. The theory is described by the parity-invariant
action:
    \bea
     \label{a}I_{BHT}=\frac{1}{16\pi G}\int d^3x
     \sqrt{-g}[R-2\lambda-\frac{1}{m^2}(R_{\mu\nu}R^{\mu\nu}-\frac{3}{8}
      R^2)]
     \eea
This action yields fourth order field equations for the metric. It is
known that, for a special case where $m^2=\lambda$, the theory has a
unique maximal solution with interesting features such as
enhancement of gauge invariance for the linearized theory. It is
also shown that, in this same case, the Brown-Henneaux boundary
conditions can be consistently relaxed, which enlarges the space
of admissible solutions so as to include rotating black holes,
gravitational solitons, kinks and wormholes \cite{BOOST,J}.

Some explicit expressions have been computed for the corrected entropy of such
black holes as Lovelock  and BTZ \cite{BTZ1,trace,lov,kerr,BTZ2}. In this paper, we calculate the
corrected temperature and entropy for an asymptotically AdS
rotating black hole solution of the new massive gravity
\cite{OLIVA2}.

The outline of this paper is  as follows: In Section $2$, we will
consider the field equations of the black hole solution  for the
new massive gravity theory at the horizons  express in the
first law of thermodynamics. In Section $3$, the thermodynamic
properties of a rotating black hole solution are summarized and
the semiclassical entropy and temperature are written. In Section
$4$, the Cardy formula is used to compute the first-order quantum
correction to the Bekenstein-Hawking entropy. In
Sections $5$ and $6$, the corrected  Hawking temperature and
entropy are calculated using the tunneling method. In Section
$7$, an unknown parameter $\beta_1$ in the tunneling method is
identified.
\section{Relationship between field equations and thermodynamics}
In the special case of $m^2=\lambda$ and negative cosmological
constant, $\lambda=-\frac{1}{2l^2}$, the metric of the BHT massive
gravity theory admits the following exact solution \cite{J}
    \bea
    \label{a3} ds^2=-f(r) dt^2+\frac{dr^2}{f(r)} +r^2 d\phi^2,
    \eea
where,
    \bea
    \label{a4}
    &&f(r)=\frac{r^2}{l^2}+b r-\mu,\\
    &&b=-\frac{1}{l^2}(r_+ +r_-),\\\nonumber
    &&\mu=-\frac{r_+ r_-}{l^2}.
    \eea
Now, let us consider $L_m$, Lagrangian density of matter fields, in
the action of the BHT massive gravity theory. Varying the action
yields corresponding equation of motion
    \bea
    \label{a1} G_{\mu \nu}+\lambda
    g_{\mu\nu}-\frac{1}{2m^2}K_{\mu\nu}=8\pi G T^m_{\mu\nu},
    \eea
where,
   \bea
    \label{a2}
    K_{\mu\nu}&\equiv& 2\nabla^2
    R_{\mu\nu}-\frac{1}{2}(\nabla_\mu\nabla_\nu R+
    g_{\mu\nu}\nabla^2 R )\nonumber\\
    &-&8 R_{\mu\rho} R^\rho _\nu+ \frac{9}{2}R
    R_{\mu\nu}+g_{\mu\nu}[3R^{\alpha\beta}R_{\alpha\beta}-\frac{13}{8}R^2],
    \eea
Here, $T^m_{\mu\nu}$ is the energy-momentum tensor for matter
fields. The components of the Einstein tensor at the horizons for
the metric (\ref{a3}) are given by \cite{akbar}
    \bea
    \label{a6}
    &&G_0^0|_{r=r_+}=G_1^1|_{r=r_+}=\frac{f'(r_+)}{2r_+}\\\nonumber
    &&G_0^0|_{r=r_-}=G_1^1|_{r=r_-}=\frac{f'(r_-)}{2r_-}.
    \eea
Also, the field equations (\ref{a1}) can be cast in the following form
    \bea
    \label{a7} G_{\mu \nu}+\lambda
    g_{\mu\nu}=8\pi G (T^m_{\mu\nu}+\frac{1}{8\pi G}
    T^{(cur)}_{\mu\nu}),
    \eea
where,

\bea
    \label{a777}
    T^{(cur)}_{\mu\nu}=\frac{1}{2m^2}K_{\mu\nu}.
    \eea
\noindent It is straightforward to show that  $T_0^{0 (cur)}$ and
$T_1^{1(cur)}$ are the same at the horizons: $r_+$ and $r_-$.
    \bea
    \label{a8}
    &&T_0^{0(cur)}|_{r=r_+}=T_1^{1(cur)}|_{r=r_+}=\frac{b}{2r_+}+\frac{1}{2l^2}\nonumber\\
    &&T_0^{0(cur)}|_{r=r_-}=T_1^{1(cur)}|_{r=r_-}=\frac{b}{2r_-}+\frac{1}{2l^2},
    \eea
From the above equations and the field equations for the stress-energy
tensor of matter fields at the horizons, we have
    \bea
    \label{a9}T_0^{0 (m)}=T_1^{1(m)}.
    \eea
At the horizons, the 0-0 component of the field equations can be
written as
    \bea
    \label{a10}
    \frac{f'(r_+)}{2r_+}-\frac{1}{l^2}-\frac{b}{2r_+}=8\pi G
    P_{{r}_{+}}
    \eea
and
    \bea
    \label{aa10}
    \frac{f'(r_-)}{2r_-}-\frac{1}{l^2}-\frac{b}{2r_-}=8\pi G P_{{r}_{-}},
    \eea
where, $P_r$ is the radial pressure. Consider a displacement $dr_+$
and $dr_-$ multiplied by both sides of (\ref{a10}) and
(\ref{aa10})  when they are subtracted  $[f'(r_+)=-f'(r_-)]$, we get
    \bea
    \label{a11}
    \frac{f'(r_+)}{4\pi}(\frac{\pi}{2G} d(r_+-r_-))&-&\frac{r_+-r_-}{8 G l^2}(dr_+-dr_-)\nonumber\\
    &=&Pd(V_+-V_-).
    \eea
The Hawking temperature and Bekenstein-Hawking entropy are given
by
     \bea
    \label{a5}
    &&T=\frac{r_+-r_-}{4\pi l^2}=\frac{f'(r_+)}{4\pi}\\
    &&S=\frac{A_+ -A_-}{4G},
    \eea
\noindent where, $A_{\pm}=2\pi r_{\pm}$. Thus, the first term in Eq
(\ref{a11}) represents $TdS$ and the second term can be identified
as the mass term. Finally, this equation can be written as follows
    \bea
    \label{a12}
    T dS-d(\Delta M)=Pd(V_+-V_-)
    \eea
\noindent which is same as the first law of thermodynamics.
\section{Thermodynamics of a rotating black hole}
Considering the special case ($m^2=\lambda=-\frac{1}{2l^2}$) for
the BHT theory, the rotating black hole solution is given in the
following way \cite{BOOST,J}
     \bea
    \label{1} ds^2=-NF dt^2+\frac{dr^2}{F} + r^2(d\phi +N^{\phi} dt)^2,
    \eea
here
    \bea
    \label{2}
    &&N=[1+\frac{bl^2}{4H}(1-\Xi^{\frac{1}{2}})]^2\nonumber\\
    &&N_{\Phi}=-\frac{a}{2r^2}(4GM-b H)\nonumber\\
    &&F=\frac {H^2}{r^2} [\frac
    {H^2}{l^2}+\frac{b}{2}(1+\Xi^{\frac{1}{2}})H + \frac
    {b^2l^2}{16}(1-\Xi^{\frac{1}{2}})^2-4GM\Xi^{\frac{1}{2}}],\nonumber\\
    \eea
and
    \bea
    \label{3} H=[r^2-2GMl^2(1-\Xi^{\frac{1}{2}})-\frac
    {b^2l^4}{16}(1-\Xi^{\frac{1}{2}})^2]^{\frac{1}{2}},
    \eea
where $\Xi=1-\frac{a^2}{l^2}$, the parameter $a$ is bounded by AdS
radius $-l\leq a  \leq\ l$ . This solution is explained by two
global charges, being $M$ and $J=Ma$ which are the mass and
angular momentum, respectively, as well as by an additional
``gravitational hair" parameter, $b$.

The event horizon radius and temperature can be expressed in a
more convenient form $r_+=\gamma \bar{r} _+$ and $T=\gamma^{-1}
\bar{T} $ where $\gamma^2=\frac{1}{2}(1+\Xi ^{-\frac{1}{2}})$.
Here, $\bar{r}_+$ is identified as the event horizon radius and
$\bar{T}$ as the temperature for the static case.  Angular velocity of
the horizon is given by \cite{OLIVA2}
     \bea
     \label{4} \Omega_+=\frac{1}{a}(\Xi^{\frac{1}{2}}-1).
     \eea
The semi-classical Hawking temperature can be derived as follows
     \bea
     \label{5} T=\frac{\hbar}{\pi l}\Xi^{\frac{1}{2}}\sqrt{2G\Delta
     M(1+\Xi^{\frac{1}{2}})^{-1}},
     \eea
where
      \bea
     \label{6} \Delta M:=M-M_0=M+\frac{b^2 l^2}{16G}.
     \eea
Consider the first law of thermodynamics for a chargeless
rotating black hole
     \bea
    \label{7} d(\Delta M)=T dS-\Omega_+ d(\Delta J),
    \eea
here $\Delta J= M a- M_0 a$. We can express $dS$ as a first order
differential equation
 \bea
    \label{8}
    dS&=&\frac{d(\Delta M)}{T}+\frac{\Omega_+}{T} d(\Delta
    J)\nonumber\\
    &=&(\frac{1}{T}+\frac{a\Omega_+}{T}) dM - (\frac{1}{T}+\frac{a\Omega_+}{T})
    dM_0+\frac{\Omega_+(M-M_0)}{T} da,\nonumber\\
    \eea
The angular velocity (\ref{4}) can be simplified to
    \bea
    \label{9} \Omega_+=-\frac{\Delta J}{l^2(\Delta M+\sqrt{\Delta M^2-\frac{\Delta
    J^2}{l^2}})}.
    \eea
First order partial differential equation (\ref{8}) is exact if it
fulfills the following integrability conditions \cite{BTZ}
    \bea
    \label{12}
    &&\frac{\partial}{\partial M}(\frac{1}{T}+\frac{a\Omega_+}{T})=-\frac{\partial}{\partial
    M_0}(\frac{1}{T}+\frac{a\Omega_+}{T})\nonumber\\
    &&\frac{\partial}{\partial a}(\frac{1}{T}+\frac{a\Omega_+}{T})=\frac{\partial}{\partial
    M}(\frac{\Omega_+(M-M_0)}{T})\nonumber\\
    &&\frac{\partial}{\partial a}(\frac{1}{T}+\frac{a\Omega_+}{T})=-\frac{\partial}{\partial
    M_0}(\frac{\Omega_+(M-M_0)}{T}).
    \eea
Using these three conditions, the solution of (\ref{8}) is given by
    \bea
    \label{16}
    S=\frac{\pi l}{\hbar}\sqrt{\frac{2}{G}\Delta
    M(1+\Xi^{\frac{1}{2}})},
    \eea
For $b=0$, the  entropy and temperature in these solutions reduce
to the BTZ black hole \cite{BTZ1,BTZ2}.
\section{Correction to the semiclassical entropy using the Cardy formula }
The Bekenstein-Hawking entropy of a
black hole may be computed by counting microscopic states and using the Cardy formula \cite{7}. The density
of states could be determined by contour integration from the
partition function. The partition function is given by
   \bea
    \label{18} Z(\tau,\bar{\tau})=\textrm{Tr} e^{2\pi i\tau L_0} e^{-2\pi i\bar{\tau} \bar{L}_0}=\sum \rho(\Delta,\bar{\Delta}) e^{2\pi i\Delta\tau} e^{-2\pi
    i\bar{\Delta}\bar{\tau}}.
    \eea
For a unitary theory, $\rho$ is the number of states with
eigenvalues $L_0=\Delta$ and $\bar{L}_0=\bar{\Delta}$, as witnessed by inserting a complete set of states into the trace. For a
nonunitary theory, $\rho$ is the difference between the number of
positive- and negative-norm states with appropriate eigenvalues.
Consider $q=e^{2\pi i\tau}$ and $\bar{q}=e^{2\pi i\bar{\tau}}$, we
get
    \bea
    \label{19} \rho(\Delta,\bar{\Delta})=\frac{1}{(2\pi i)^2} \int
    \frac{dq}{q^{\Delta+1}}\frac{d\bar{q}}{\bar{q}^{\bar{\Delta}+1}}Z(q,\bar{q}),
    \eea
One can easily find $\rho(\Delta)$ in this way
    \bea
    \label{20} \rho(\Delta)\thickapprox (\frac{c}{96 \Delta^3})^{\frac{1}{4}}
    \exp{2\pi\sqrt{\frac{c\Delta}{6}}}.
    \eea
As shown in \cite{J,OLIVA2}, the algebra of the conserved charges
also acquires a central extension being twice the value found for
general relativity
    \bea
    \label{21} c=\bar{c}=\frac{3l}{G}.
    \eea
Now, the generators of the Brown-Henneaux Virasoro algebras can be
computed explicitly, simply as the Hamiltonian and momentum
constraints of general relativity smeared against appropriate
vector fields, for the rotating black hole
    \bea
    \label{22}
    &&\Delta=\frac{1}{2}(l\Delta M+\Delta J)=\frac{1}{2}\Delta M(l+a)\\\nonumber
    &&\bar{\Delta}=\frac{1}{2}(l\Delta M-\Delta J)=\frac{1}{2}\Delta M(l-a).
    \eea
Using (\ref{21}) and (\ref{22}), we have
     \bea
    \label{23} 2\pi\sqrt{\frac{c\Delta}{6}}+2\pi\sqrt{\frac{\bar{c}\bar{\Delta}}{6}}=\pi l\sqrt{\frac{2}{G}\Delta M(1+\Xi^{\frac{1}{2}})},
    \eea
which is exactly equal to the  semiclassical entropy. To obtain
the logarithmic correction to the entropy, we calculate the density of
states from (\ref{20})
    \bea
    \label{24}
    \rho(\Delta,\bar{\Delta})\approx (\frac{l}{4G((l \Delta M)^2-(\Delta J)^2)^{\frac{3}{2}}})^{\frac{1}{2}}  \exp(\pi l\sqrt{\frac{2}{G}\Delta
    M(1+\Xi^{\frac{1}{2}})})\nonumber\\
    \approx (\frac{l}{4Gl^3 (\Delta M)^3(1-\frac{a^2}{l^2})^{\frac{3}{2}}})^{\frac{1}{2}}  \exp(\pi l\sqrt{\frac{2}{G}\Delta M(1+\Xi^{\frac{1}{2}})}).\nonumber\\
    \eea
Hence, the corrected entropy is given by
     \bea
    \label{25}
    S &\sim& \pi l\sqrt{\frac{2}{G}\Delta M(1+\Xi^{\frac{1}{2}})}\nonumber\\
    &-&\frac{3}{2}\ln(\pi l\sqrt{\frac{2}{G}\Delta
    M(1+\Xi^{\frac{1}{2}})})-\frac{3}{2}\ln\kappa
    l+ const,
    \eea
where,
    \bea
    \label{26}
    \kappa=\frac{2}{l}\Xi^{\frac{1}{2}}\sqrt{\frac{2G\Delta M}{1+\Xi^{\frac{1}{2}}}}
    \eea
is the surface gravity.
\section{Corrections to the semiclassical Hawking temperature }
In this section, we consider the scalar particle by using tunneling
 method beyond semiclassical approximation for finding the correction to the
semiclassical Hawking temperature for an asymptotically AdS rotating
black hole. For this purpose, we should
isolate the $r-t$ sector of the metric from the angular part. A
coordinate transformation near the  horizon approximation can be found
    \bea
    \label{b1} d\chi=d\phi+\Omega_+ dt,
    \eea
Also the metric (\ref{2}) takes the form
    \bea
    \label{b2} ds^2=-NF dt^2+\frac{dr^2}{F} + r_+^2 d\chi^2,
    \eea
where, the $r-t$ sector is isolated from the angular part. The
massless particle in this spacetime obeys the following
Klein-Gordon equation
     \bea
     \label{b3} -\frac{\hbar ^2}{\sqrt{-g}}\partial_{\mu}[g^{\mu\nu}\sqrt{-g}\partial _{\nu}]\phi=0.
    \eea
To solve Eq (\ref{b3}) with the background metric (\ref{b2}), we
can write the standard WKB anstaz for $\phi$ as
     \bea
     \label{b5} \phi(r,t)=exp[-\frac{i}{\hbar}S(r,t)],
    \eea
Substituting the relation (\ref{b5}) in  (\ref{b3}), we get
     \bea
     \label{b4}
     \frac{1}{NF} (\frac{\partial S}{\partial t})^2&-&F(\frac{\partial S}{\partial r})^2-\frac{\hbar}{i}\frac{1}{NF}(\frac{\partial^2 S}{\partial t^2})+\frac{\hbar}{i}F(\frac{\partial^2 S}{\partial
     r^2})\nonumber\\
     &+&\frac{\hbar}{i}\partial_r F (\frac{\partial S}{\partial t})+\frac{\hbar}{i}\frac{\partial_r\sqrt{N}}{\sqrt{N}} F(\frac{\partial S}{\partial r})=0,
    \eea
In this case, $S(r, t)$ can be expanded in power of $\hbar$
   \bea
   \label{b6} S(r,t)=S_0(r,t)+\sum_i \hbar^{i} S_i(r,t).
   \eea
Putting $S(r, t)$ from  (\ref{b6}) in (\ref{b4}) will yield the following set
of equations
     \bea
     \label{b7}
     &&\hbar^0:\frac{\partial S_0}{\partial t}=\pm \sqrt{N}F \frac{\partial S_0}{\partial r}\\\nonumber
     &&\hbar^1:\frac{\partial S_1}{\partial t}=\pm \sqrt{N}F \frac{\partial S_1}{\partial r}\\\nonumber
     &&\hbar^2:\frac{\partial S_2}{\partial t}=\pm \sqrt{N}F \frac{\partial S_2}{\partial r}\\\nonumber
     &&.\\\nonumber
     &&.\\\nonumber
     &&.
    \eea
All the  equations are identical and their solutions are related to each
other. We assume that any $S_i(r, t)$ can differ from $S_0(r, t)$
by a proportionality factor. The general form of the action is given by
  \bea
  \label{b8}  S(r,t)=(1+\sum_i \gamma_i\hbar^i) S_0(r,t),
  \eea
The dimension of $\gamma _i$ is equal to the dimension of $\hbar
^{-i}$; thus, $\gamma _i$ can be expressed in terms of dimensionless
constant. In $(2+1)$ dimensions, $\hbar$  can be replaced by Planck
length$ (l_P)$, the length parameters for these black holes are $r_+ $
and $r_-$. We have
     \bea
     \label{b9} S(r,t)=(1+\sum_i \frac{\beta_i\hbar^i}{(a_1r_+ +a_2r_-)^i}) S_0(r,t),
     \eea
where $\beta _i$'s are dimensionless constants and $a_1$ and
$a_2$ will be calculated in a later stage. We can isolate the semiclassical
action for the $ r - t$ sector near the horizon
     \bea
     \label{b11} S_0(r,t)=\omega t+\tilde{S}_0(r),
     \eea
The total energy of the tunneling particle near the horizon approximation
is given by
     \bea
     \label{b12} \omega=E-J\Omega_+.
     \eea
Substituting (\ref{b11}) in the first equation of (\ref{b7}) yields,
     \bea
     \label{b13}\tilde{S}_0=\pm \omega \int\frac{dr}{\sqrt{N}F},
     \eea
the $+ (-)$ sign shows that the particle is ingoing (outgoing). As a
result of substituting  $S_0(r, t)$ from (\ref{b11}) and
$\tilde{S}_0$ from (\ref{b13}), one is able to find (\ref{b9}) as
follows
     \bea
     \label{b14} S(r,t)=(1+\sum_i \frac{\beta_i\hbar^i}{(a_1r_+ +a_2r_-)^i})(\omega \ t\pm\omega\int \frac{dr}{\sqrt{N}F}).
     \eea
A solution for the scalar field in the presence of the higher order
corrections to the semiclassical action is given by
   \begin{widetext}
    \bea
     \label{b15} \phi_{in}=\exp[-\frac{i}{\hbar}(1+\sum_i\beta_i\frac{\hbar^i}{(a_1r_+ +a_2r_-)^i})(\omega \ t+\omega\int\frac{dr}{\sqrt{N}F})]\\
     \label{b16} \phi_{out}=\exp [-\frac{i}{\hbar}(1+\sum_i\beta_i\frac{\hbar^i}{(a_1r_+ +a_2r_-)^i})(\omega \ t-\omega\int\frac{dr}{\sqrt{N}F})].
     \eea
     \end{widetext}
where ingoing and outgoing particles cross the event horizon on
different paths. Since the metric coefficients for $r - t$ sector
change sign at the two sides of the event horizon, the outgoing
particle cannot cross the event horizon classically. Therefore,
the path on which tunneling takes place has an imaginary time
coordinate ($\textrm{Im} t$). We can write the ingoing and
outgoing probabilities as
   \begin{widetext}
    \bea
    \label{b17}P_{in}=|\phi_{in}|^2=\exp [\frac{2}{\hbar}(1+\sum_i\beta_i\frac{\hbar^i}{(a_1r_+ +a_2r_-)^i})(\omega \textrm{Im} \ t+\omega \textrm{Im}\int\frac{dr}{\sqrt{N}F})],\\
     \label{b18}P_{out}=|\phi_{out}|^2=\exp [\frac{2}{\hbar}(1+\sum_i\beta_i\frac{\hbar^i}{(a_1r_+ +a_2r_-)^i})(\omega
     \textrm{Im} \ t-\omega \textrm{Im} \int\frac{dr}{\sqrt{N}F})].
     \eea
     \end{widetext}
In the classical limit, the ingoing particle probability is unity;
therefore, we have
    \bea
     \label{b19} \textrm{Im} \ t=-\textrm{Im} \int\frac{dr}{\sqrt{N}F},
     \eea
Now $P_{out}$ can be expressed as
     \bea
     \label{b20} P_{out}=\exp [-\frac{4}{\hbar}\omega(1+\sum_i\beta_i\frac{\hbar^i}{(a_1r_+ +a_2r_-)^i})
     \textrm{Im} \int\frac{dr}{\sqrt{N}F}],\nonumber\\
     \eea
We identify the temperature of this black hole by using the
principle of detailed balance for the ingoing and outgoing
probabilities
      \bea
     \label{b21} \frac{P_{out}}{P_{in}}=\exp(-\frac{\omega}{T_H}).
     \eea
The paths for the ingoing and outgoing particles crossing the event
horizon are not the same and the outgoing particle cannot cross
the event horizon classically. Considering the ingoing probability to
be unitary, we can write
     \bea
     \label{b22} P_{out}=\exp(-\frac{\omega}{T_H}).
     \eea
As a result, the corrected Hawking temperature becomes
     \bea
     \label{b23} T_H=T(1+\sum_i\beta_i\frac{\hbar^i}{(a_1r_+
     +a_2r_-)^i})^{-1},
     \eea
where, $T$ is the semiclassical Hawking temperature and other terms
are corrections to the higher order quantum effects.
\section{Entropy Correction}
Here, we are able to calculate the corrected entropy by using the
expression of the modified temperature computed in (\ref{b23})
    \bea
     \label{c1} T_H=T(1+\sum_i\beta_i\frac{\hbar^i}{(a_1r_+
     +a_2r_-)^i})^{-1}.
     \eea
Including modified Hawking temperature in the above equation, we find
a similar expression for the  thermodynamics law (\ref{8})
    \bea
    \label{c2}
    &&dS_{bh}=\frac{d(\Delta M)}{T_H}+\frac{\Omega_+}{T_H} d(\Delta
    J)\nonumber\\
    &&=(\frac{1}{T_H}+\frac{a\Omega_+}{T_H}) dM - (\frac{1}{T_H}+\frac{a\Omega_+}{T_H})dM_0+\frac{\Omega_+(M-M_0)}{T_H}
    da,\nonumber\\
    \eea
Eq. (\ref{c2}) is an exact first order differential equation if
it fulfills  the following  integrability conditions
    \bea
     \label{c3}
     &&\frac{\partial}{\partial M}(\frac{1}{T_H}+\frac{a\Omega_+}{T_H})=-\frac{\partial}{\partial
     M_0}(\frac{1}{T_H}+\frac{a\Omega_+}{T_H})\nonumber\\
     &&\frac{\partial}{\partial a}(\frac{1}{T_H}+\frac{a\Omega_+}{T_H})=\frac{\partial}{\partial
     M}(\frac{\Omega_+(M-M_0)}{T_H})\nonumber\\
     &&\frac{\partial}{\partial a}(\frac{1}{T_H}+\frac{a\Omega_+}{T_H})=-\frac{\partial}{\partial
     M_0}(\frac{\Omega_+(M-M_0)}{T_H}),
     \eea
By substituting $ \frac{\Omega_+}{T_H}$ and $\frac{1}{T_H}$ from
their definitions from (\ref{9}) and (\ref{c1}) in (\ref{c3}),  we
find $a_1=1, a_2=-1$. Now, we are able to continue the procedure
 applied for the semiclassical entropy by setting up a new
dictionary and using the functional forms in which all semiclassical
quantities are replaced by their corrected forms
        \bea
        \label{c4}
        S_{bh}=\int (\frac{1}{T}+\frac{a\Omega_+}{T})\sum_i(1+\beta_i\frac{\hbar^i}{(r_+ -r_-)^i}) dM.
        \eea
The quantum corrections up to the second order corrections in Eq
(\ref{c4}) can be written as
      \bea
      \label{c5}
      &&S_{bh}=\int (\frac{1}{T}+\frac{a\Omega_+}{T})[1+\beta_1\frac{\hbar}{(r_+ -r_-)}+\beta_2\frac{\hbar^2}{(r_+
      -r_-)^2}\nonumber\\
      &&+O(\hbar^3)]\ d M.
     \eea
For the corrected entropy, we get
      \bea
      \label{c6}
      &&S_{bh}=\frac{\pi l}{\hbar}\sqrt{\frac{2}{G}(\Delta M+\sqrt{\Delta M^2-\frac{\Delta
      J^2}{l^2}})}\nonumber\\
      &&+\beta_1\frac{\pi}{2G}\log[\frac{\pi l}{\hbar}\sqrt{\frac{2}{G}(\Delta M+\sqrt{\Delta M^2-\frac{\Delta
      J^2}{l^2}})}]\nonumber\\
      &&-\beta_2\frac{\pi^2 }{4 G^2}\frac{\hbar}{\pi l\sqrt{\frac{2}{G}(\Delta M+\sqrt{\Delta M^2-\frac{\Delta
      J^2}{l^2}})}}\nonumber\\
      &&+O(\hbar^3).
      \eea

Thus the corrected entropy can be written in terms of
semiclassical entropy
       \bea
      \label{c7}
      &&S_{bh}=S_{cl}+ \beta_1\frac{\pi}{2G}\log(S_{cl})-\beta_2\frac{\pi^2}{4
      G^2}\frac{1}{S_{cl}}+O(\hbar^3).
      \eea
In this equation, the first sentence is the semiclassical entropy that,
in the limit $b=0$, is equal to the entropy of BTZ black hole. The
other terms are related to the quantum corrections.

\noindent A standard formalism to identify the first coefficient
of the leading correction is trace anomaly \cite{trace,lov,kerr}.
However, by comparing (\ref{25}) and (\ref{c7}), we have
$\beta_1=-\frac{3G}{\pi}$.

It should be noted that the semiclassical entropy $S_{cl}$ can be
calculated by Wald's formula \cite{J,wald}. Wald's formula calculates
 the entropy of the black hole when the
action contains terms of higher order in the curvature tensor.
This formula mentions the entropy as an integral over the horizon
of the black hole in the following way,
      \bea
      \label{c7}
      S=-2\pi \int_{\Sigma_{h}} \frac{\partial L}{\partial R_{\mu\nu\rho\lambda}}\epsilon_{\mu\nu} \epsilon_{\rho\lambda}\bar{\epsilon} .
      \eea

\noindent \noindent where $L$ is the lagrangian, $\epsilon_{\mu\nu}$ is the binormal vector to the space-like bifurcation surface $\Sigma_{h}$ and $\bar{\epsilon}$ represents the volume form \cite{jacob}. However, it is usually  not simple to relate Wald's formula to
the logarithmic correction that emerges from different theories related to quantum gravity
such as loop gravity, entanglement entropy, tunneling method etc \cite{Aros,Carav}.
For obtaining quantum corrections to the entropy by Wald's formula we need to know full effective action.
Tunneling method provides us with a corrected entropy and temperature which can also be applied to
black holes with the higher derivative terms such as Lovelock black holes \cite{lov}. We do not know a systematic
method to produce higher derivative terms in the Lagrangian and then use Wald's formula to get entropy corrections.
This is an interesting problem and its solution may sheds light on the universality problem that exist in different theories of quantum gravity \cite{Carlip}.

\section{Conclusion}
We express the field equations of AdS black hole solution of the
new  massive gravity as the first law of thermodynamics. The corrected
semi-classical entropy of the asymptotically AdS rotating black hole
of the new massive gravity is calculated  by using the cardy
formula and tunneling formalism. It is found that the leading
correction to the semiclassical entropy for black hole is
logarithmic and next to the leading correction is the  inverse of the
semiclassical entropy. The  two methods yield the same results. The
leading correction coefficient can be obtained by using the Cardy
formula. The new massive gravity is expected to be unitary and
renormalizable. It must, therefore, be interesting to obtain these results
through a more direct use of the field theory of the new
massive  theory.


\end{document}